\documentclass[aip,cha,10pt,longbibliography,unsortedaddress,floatfix]{revtex4-2}

\usepackage{amsmath,amssymb,amsfonts,amsthm}
\usepackage{mathtools}
\usepackage{bm}

\usepackage{graphicx}
\usepackage{color}

\usepackage[T1]{fontenc}
\usepackage{mathptmx}

\newcommand{\BF}{\bm{F}}
\newcommand{\BG}{\bm{G}}
\newcommand{\BJ}{\bm{J}}
\newcommand{\BX}{\bm{X}}
\newcommand{\BZ}{\bm{Z}}
\newcommand{\Bf}{\bm{f}}
\newcommand{\Bp}{\bm{p}}
\newcommand{\calB}{\mathcal{B}}
\newcommand{\calN}{\mathcal{N}}
\newcommand{\Bchi}{\boldsymbol{\chi}}

\newcommand{\inner}[2]{#1 \cdot #2}
\newcommand{\average}[2]{\left\langle #1 \right\rangle_{#2}}

\newcommand{\norm}[1]{\left\| #1 \right\|}
\newcommand{\fd}[2]{\frac{\delta #1}{\delta #2}}

\begin{document}
\title{Optimal interaction functions realizing higher-order Kuramoto dynamics with arbitrary limit-cycle oscillators}
\author{Norihisa Namura}
\thanks{Corresponding author. E-mail: namura.n.aa@m.titech.ac.jp}
\affiliation{Department of Systems and Control Engineering, Institute of Science Tokyo, Tokyo, Japan}
\author{Riccardo Muolo}
\affiliation{Department of Systems and Control Engineering, Institute of Science Tokyo, Tokyo, Japan}
\author{Hiroya Nakao}
\affiliation{Department of Systems and Control Engineering, Institute of Science Tokyo, Tokyo, Japan}
\affiliation{Research Center for Autonomous Systems Materialogy, Institute of Science Tokyo, Yokohama, Japan}

\date{\today}


\begin{abstract}

The Kuramoto model is the simplest case of globally coupled phase oscillators with a purely sinusoidal fundamental-harmonic phase coupling function, whose dynamical properties have been extensively studied.
While coupled phase oscillators are derived from weakly interacting limit-cycle oscillators via phase reduction, this procedure does not necessarily yield the Kuramoto model or its higher-order extensions exactly for general limit-cycle oscillators and interaction functions, except in the special case of interacting Stuart--Landau oscillators.
In this study, we artificially design optimal pairwise and higher-order interaction functions between limit-cycle oscillators, from which higher-order Kuramoto models can be exactly derived via phase reduction for arbitrary smooth limit-cycle oscillators.
We validate the results through numerical simulations of FitzHugh--Nagumo oscillators, demonstrating that the collective synchronization dynamics predicted by the reduced higher-order Kuramoto models are realized. 
Control of the collective phase of the FitzHugh--Nagumo oscillators based on Ott-Antonsen reduction of the higher-order Kuramoto model is also demonstrated.
\end{abstract}

\maketitle


\section*{Lead Paragraph}

In complex systems, the collective behavior is shaped by the nature of the interactions between elementary components. 
One prominent example is the collective synchronization of rhythmic elements, as captured by the classical Kuramoto model, where the pairwise coupling between the oscillators affects the transition from incoherence to synchronization.
Recent generalizations of the Kuramoto model to higher-order interactions yield even richer dynamics that are not observed with purely pairwise coupling.
The Kuramoto model is derived via phase reduction from coupled Stuart--Landau oscillators, which represent the normal form of a supercritical Hopf bifurcation with rotationally symmetric dynamics. 
It has been extensively studied, and many important properties have been revealed.
However, for general limit-cycle oscillators and interaction functions, phase reduction does not necessarily yield the Kuramoto model but instead gives more complex phase models whose phase coupling functions possess higher-harmonic components.
In this study, we propose a special class of interaction functions between general limit-cycle oscillators that exactly allow us to obtain the higher-order Kuramoto model via phase reduction.
The proposed method enables arbitrary systems of interacting limit-cycle oscillators to behave like the simplest higher-order Kuramoto models, facilitating the design and control of their collective dynamics.


\section{Introduction}

Self-sustained rhythms can be mathematically modeled as limit-cycle oscillators, which capture the essential features of systems with stable periodic behavior~\cite{winfree1967biological}, such as electric circuits~\cite{van1926relaxation}, neurons~\cite{Hodgkin1952quantitative}, or chemical oscillators~\cite{Kuramoto1984chemical,Strogatz2015nonlinear}. 
When such oscillators interact, mutual synchronization, in which the oscillator phases tend to align with each other, can emerge~\cite{winfree1967biological}. 
Synchronization phenomena are a typical example of self-organization in complex systems and are ubiquitous in natural and engineered systems~\cite{Pikovsky2001synchronization,Strogatz2003sync,arenas2008synchronization,boccaletti2018synchronization}. 
Real-world examples include power grids~\cite{Dorfler2012synchronization}, gait generation in legged robots~\cite{Namura2025central}, heartbeats and respiration, fireflies' blinking, frogs' croaking, and bridge vibrations~\cite{Strogatz2003sync}.
Such collective rhythmic behavior arises from the interactions among individual self-sustained oscillators. 
Understanding and controlling these collective rhythmic behaviors is a central objective in many fields, including both nonlinear physics and control theory~\cite{d2023controlling}.

For analyzing mutual synchronization, phase reduction theory is useful since it reduces a high-dimensional interacting oscillatory system to a simple system of coupled phase oscillators governed by their intrinsic frequency and phase coupling functions~\cite{Kuramoto1984chemical,Winfree2001geometry,Brown2004phase,Ermentrout2010mathematical,Nakao2016phase,Monga2019phase1,pietras2019network}.
One of the most important results of phase reduction\footnote{To be precise, the original derivation of the Kuramoto model in 1975 was achieved through an adiabatic elimination, while successively the phase reduction method was applied.} is the Kuramoto model~\cite{Kuramoto1975self,Kuramoto1984chemical,strogatz2000kuramoto,acebron2005kuramoto}, originally derived from globally coupled Stuart--Landau oscillators. 
As the coupling strength increases, the system undergoes a dynamical transition to collective synchronization~\cite{Kuramoto1984chemical,Chiba2015proof}.
The original Kuramoto model assumes that the coupling is global (all-to-all), pairwise, and given by a purely sinusoidal function of the fundamental harmonic.
The simplicity of the Kuramoto model has facilitated various analytical approaches, which provide deep insights into the collective synchronization transitions. 
However, when phase reduction is applied to general limit-cycle oscillators and interaction functions describing real-world systems, 
the resulting phase coupling functions generally possess higher-harmonic components, leading to more complex phase models that can be difficult to analyze or control. 

Recently, higher-order (many-body) interactions via hypergraphs or simplicial complexes are attracting much attention as extensions of networks~\cite{battiston2020networks,bianconi2021higher,natphys,bick2023higher,boccaletti2023structure,muolo2024turing,millan2025topology}. 
Phase oscillators with higher-order interactions have also been studied actively, revealing their nontrivial dynamics that cannot be observed in the original Kuramoto model~\cite{millan2020explosive,gambuzza2021stability,gallo2022synchronization,della2023emergence,tanaka2011multistable,Skardal2020higher,lucas2020multiorder}.
It has been shown that, for limit-cycle oscillators with pairwise~(two-body) interactions, the second-order phase reduction leads to phase equation with higher-order~(three-body) coupling~\cite{ashwin2016hopf,Leon2019phase,gengel2020high,bick2024higher,mau2024phase}, while the ordinary (first-order) phase reduction yields phase oscillators with pairwise coupling.
In the case of second-order phase reduction, the resulting higher-order coupling can affect qualitatively the collective dynamics in the original system.
Alternatively, we may also consider limit-cycle oscillators with higher-order interactions and apply the first-order phase reduction to obtain coupled phase oscillators~\cite{Leon2024higher,Leon2025theory}. 
In the aforementioned cases, if we apply phase reduction to general oscillators and interaction functions, the resulting phase coupling functions are not purely sinusoidal of the fundamental harmonic but also contain high-harmonic components, leading to difficulty in the analysis.

In this study, extending the work for optimizing coupling functions~\cite{Namura2024optimal}, we artificially design optimal interaction functions that exactly yield the pairwise and two types of higher-order Kuramoto models from any system of nearly identical limit-cycle oscillators by phase reduction, without any approximation in the phase coupling functions~\cite{Daido1996onset}.
The optimal interaction functions are derived so that they minimize the averaged power of the interaction function while realizing a desired phase coupling function.
To validate the proposed interaction functions, we perform numerical simulations for a system of interacting FitzHugh--Nagumo~(FHN) oscillators~\cite{Fitzhugh1961impulses,Nagumo1962Active}. 
We first demonstrate that, in the case only with pairwise interactions, the system exhibits a collective synchronization transition as predicted by the pairwise Kuramoto model.
Then, we show that, as predicted by the analysis of higher-order Kuramoto model with three-body interactions by Le\'{o}n \textit{et al.}~\cite{Leon2024higher}, the system undergoes an anomalous transition to synchrony, exhibiting slow switching and two-cluster states before achieving full synchronization.
We further compare the relative phase distributions between the heterogeneous system of FHN oscillators and the higher-order Kuramoto model in the two-cluster state, showing good agreement.
Finally, we realize another system of FHN oscillators that behave like the higher-order Kuramoto model amenable to Ott-Antonsen ansatz~\cite{Ott2008low,Skardal2020higher} and control its collective dynamics~\cite{Fujii2025optimal}.

This paper is organized as follows.
We introduce the optimal interaction functions and derive the higher-order Kuramoto models in Sec.~\ref{sec:model}.
Analyses of the higher-order Kuramoto models are given in Sec.~\ref{sec:analysis}.
In Sec.~\ref{sec:results}, we demonstrate the numerical results using FitzHugh--Nagumo oscillators as an example.
Discussions and conclusions are described in Sec.~\ref{sec:conclusions}.
Appendix~\ref{sec:appendix} provides the derivation of the optimal interaction functions.


\section{Derivation of the higher-order Kuramoto model}
\label{sec:model}

\subsection{Phase reduction}

We consider a system of $N$ limit-cycle oscillators with nearly identical properties that have pairwise and three-body all-to-all interactions.
The dynamics of the $j$th oscillator $(j = 1,2,\dots,N)$ is described by
\begin{align}
\label{eq:oscillator_model}
\dot{\BX}_{j} ={}& \BF_{j}(\BX_{j}) + \frac{K_{1}}{N} \sum_{k=1}^{N} \tilde{\BG}_{1}(\BX_{j}, \BX_{k})
+ \frac{K_{2}}{N^{2}} \sum_{k,\ell=1}^{N} \tilde{\BG}_{2}(\BX_{j}, \BX_{k}, \BX_{\ell}),
\end{align}
where $\BX_{j}(t) \in \mathbb{R}^{M}$ is an $M$-dimensional state of the oscillator,
$t$ is the time, 
the overdot represents the time derivative, and
$\BF_{j}: \mathbb{R}^{M} \to \mathbb{R}^{M}$ is a sufficiently smooth function representing the vector field of the $j$th oscillator.
In this study, we consider pairwise (two-body) interactions, whose effects are modeled by a sufficiently smooth interaction function $\tilde{\BG}_{1}: \mathbb{R}^{M} \times \mathbb{R}^{M} \to \mathbb{R}^{M}$, and
three-body interactions, modeled by a sufficiently smooth interaction function $\tilde{\BG}_{2}: \mathbb{R}^{M} \times \mathbb{R}^{M} \times \mathbb{R}^{M} \to \mathbb{R}^{M}$; $K_{1}$ and $K_{2}$ are the strengths of the pairwise and three-body interactions, respectively.
We assume that $\tilde{\BG}_{1}$, $\tilde{\BG}_{2}$ are of order $1$ and $K_{1}$, $K_{2}$ are of order $\epsilon$ ($0 \leq \varepsilon \ll 1$), where we consider the interaction terms as perturbations to the oscillator.
Note that what follows can be straightforwardly generalized to $d$-body interactions ($d \geq 4$).

First, we derive a system of coupled phase equations from Eq.~\eqref{eq:oscillator_model} using the first-order phase reduction.
We consider the case that the properties of the $N$ oscillators are nearly identical, i.e., we can separate the vector field $\BF_{j}$ of each oscillator into a common part $\BF$ and a small deviation $\Bf_{j}$ as
\begin{align}
\BF_{j}(\BX) = \BF(\BX) + \Bf_{j}(\BX).
\end{align}
We then assume that the common part $\BF$ has an exponentially stable limit cycle with a period $T$ and natural frequency $\omega_{0} = 2\pi/T$, 
whose orbit is represented by a $T$-periodic function $\tilde{\Bchi}(t)$ satisfying $\tilde{\Bchi}(t + T) = \tilde{\Bchi}(t)$.

Let us now define the phase with respect to the limit cycle. 
We introduce the asymptotic phase function $\Theta: \calB \to [0, 2\pi)$ in the basin of the limit cycle $\calB \subseteq \mathbb{R}^{M}$, 
so that $\inner{\nabla\Theta(\BX)}{\BF(\BX)} = \omega_{0}$ holds for any state $\BX \in \calB$,
where $\inner{\bm{a}}{\bm{b}} = \sum_{m=1}^{M} a_{m} b_{m}$ represents the scalar product of two vectors $\bm{a}$, $\bm{b} \in \mathbb{R}^{M}$ and $\nabla$ represents the gradient. 
With this formulation, the phase defined as $\theta = \Theta(\BX)$ increases with the constant frequency $\omega_{0}$, i.e.,
\begin{align}
\dot{\theta}(t) = \inner{\nabla\Theta(\BX(t))}{\BF(\BX(t))} = \omega_{0}.
\end{align}
Due to the periodicity, $\theta = 0$ and $\theta = 2\pi$ are identical.
We can represent a state on the limit cycle by $\Bchi(\theta) = \tilde{\Bchi}(t = \theta/\omega_{0})$ as a function of $\theta$,
which is a $2\pi$-periodic function satisfying $\Bchi(\theta) = \Bchi(\theta + 2\pi)$.

We now define the phase of each oscillator as $\theta_{j} = \Theta(\BX_{j})$.
We consider that the small deviation $\Bf_{j}(\BX)$, pairwise interaction terms, and three-body interaction terms as weak perturbations.
Assuming that the initial states of all oscillators are sufficiently close to the limit cycle, i.e., $\BX_{j}(0) = \Bchi(\theta_{j}(0)) + O(\varepsilon)$, and the perturbations to the common part $\BF$,
from both the deviation $\Bf_{j}$ and the interaction terms, are also of order $\varepsilon$, 
we can consider that all the trajectories remain sufficiently close to the limit cycle, i.e., $\BX_{j}(t) = \Bchi(\theta_{j}(t)) + O(\varepsilon)$ for all $t$.
Therefore, the dynamics of each phase $\theta_{j}$ can be captured with a good approximation by the following phase model obtained through phase reduction~\cite{Nakao2016phase} from Eq.~\eqref{eq:oscillator_model}, i.e.,
\begin{align}
\dot{\theta}_{j} ={}& \inner{\nabla\Theta(\BX_{j})}{\left( \BF(\BX_{j}) + \Bf_{j}(\BX_{j}) + \frac{K_{1}}{N} \sum_{k=1}^{N} \tilde{\BG}_{1}(\BX_{j}, \BX_{k}) + \frac{K_{2}}{N^{2}} \sum_{k,\ell=1}^{N} \tilde{\BG}_{2}(\BX_{j}, \BX_{k}, \BX_{\ell}) \right)} \cr
={}& \omega_{0} + \inner{\BZ(\theta_{j})}{\left( \Bf_{j}(\Bchi(\theta_{j})) + \frac{K_{1}}{N} \sum_{k=1}^{N} \BG_{1}(\theta_{j}, \theta_{k}) + \frac{K_{2}}{N^{2}} \sum_{k,\ell=1}^{N} \BG_{2}(\theta_{j}, \theta_{k}, \theta_{\ell}) \right)}
\end{align}
up to order $\varepsilon$.
In the above equation, we defined
\begin{align}
\BG_{1}(\theta_{j}, \theta_{k}) &= \tilde{\BG}_{1}(\Bchi (\theta_{j}), \Bchi(\theta_{k}))
\end{align}
and
\begin{align}
\BG_{2}(\theta_{j}, \theta_{k}, \theta_{\ell}) &= \tilde{\BG}_{2}(\Bchi(\theta_{j}), \Bchi(\theta_{k}), \Bchi(\theta_{\ell})),
\end{align}
respectively, which we call interaction functions as functions of phases (IFPs) in this study.
The function $\BZ$ is the \textit{phase sensitivity function}~(PSF)~\cite{Kuramoto1984chemical,Nakao2016phase},
also known as the infinitesimal phase response curve~(iPRC)~\cite{Ermentrout2010mathematical}, defined by $\BZ(\theta) = \nabla \Theta(\BX)|_{\BX = \Bchi(\theta)}$.
The PSF characterizes the linear phase response of the oscillator state on the limit cycle to weak perturbations,
and can be obtained as a $2\pi$-periodic solution to the \textit{adjoint equation}~\cite{Brown2004phase,Ermentrout2010mathematical,Nakao2016phase}:
\begin{align}
\omega_{0} \frac{d}{d\theta}\BZ(\theta) = -\BJ(\Bchi(\theta))^{\top}\BZ(\theta).
\end{align}
Here, ``$\top$'' represents the matrix transposition and $\BJ(\BX)$ is the Jacobian matrix of the vector field, i.e., $\BJ(\BX) = \nabla\BF(\BX)$.
The PSF should also satisfy the normalization condition $\inner{\BZ(\theta)}{d\Bchi(\theta)/d\theta} = 1$ for all $\theta$.

Since the perturbations to the common part $\BF$ are of order $\varepsilon$, the relative phase $\phi_{j}(t) = \theta_{j}(t) - \omega_{0} t$ is a slow variable for each oscillator $j$.
Therefore, we can perform the averaging approximation~\cite{Kuramoto1984chemical,Hoppensteadt1997weakly,Nakao2016phase}
and obtain the following phase equation up to order $\varepsilon$:
\begin{align}
\dot{\theta}_{j} ={}& \omega_{j} + \frac{K_{1}}{N} \sum_{k=1}^{N} \Gamma_{1}(\theta_{j} - \theta_{k})
+ \frac{K_{2}}{N^{2}} \sum_{k,\ell=1}^{N} \Gamma_{2}(\theta_{j} - \theta_{k}, \theta_{j} - \theta_{\ell}).
\end{align}
Here, $\omega_{j}$ is the frequency of the oscillator $j$ represented as $\omega_{j} = \omega_{0} + \Delta \omega_{j}$, where $\Delta \omega_{j}$ is the averaged frequency mismatch from the natural frequency $\omega_{0}$ given by
\begin{align}
\Delta \omega_{j} = \average{\inner{\BZ(\psi)}{\Bf_{j}(\Bchi(\psi))}}{\psi}.
\end{align}
In the above expression, we defined the averaging of a smooth function $g$ over one period of oscillation by
\begin{align}
\average{g(\psi)}{\psi} = \frac{1}{2\pi}\int_{0}^{2\pi} g(\psi)d\psi.
\end{align}
The function $\Gamma_{1}$ is a pairwise \textit{phase coupling function}~(PCF), which is defined as 
\begin{align}
\Gamma_{1}(\varphi_{jk}) &= \frac{1}{T} \int_{t'}^{t' + T} \inner{\BZ(\phi_{j} + \omega_{0} t)}{\BG_{1}(\phi_{j} + \omega_{0} t, \phi_{k} + \omega_{0} t)} dt \cr
&= \frac{1}{2\pi} \int_{0}^{2\pi} \inner{\BZ(\psi)}{\BG_{1}(\psi, \psi - \varphi_{jk})} d\psi \cr
&= \average{\inner{\BZ(\psi)}{\BG_{1}(\psi, \psi - \varphi_{jk})}}{\psi},
\end{align}
and $\Gamma_{2}$ is a three-body PCF, defined as 
\begin{align}
&\Gamma_{2}(\varphi_{jk}, \varphi_{j\ell}) \cr
={}& \frac{1}{T} \int_{t'}^{t' + T} \inner{\BZ(\phi_{j} + \omega_{0} t)}{\BG_{2}(\phi_{j} + \omega_{0} t, \phi_{k} + \omega_{0} t, \phi_{\ell} + \omega_{0} t)} dt \cr
={}& \frac{1}{2\pi} \int_{0}^{2\pi} \inner{\BZ(\psi)}{\BG_{2}(\psi, \psi - \varphi_{jk}, \psi - \varphi_{j\ell})} d\psi \cr
={}& \average{\inner{\BZ(\psi)}{\BG_{2}(\psi, \psi - \varphi_{jk}, \psi - \varphi_{j\ell})}}{\psi},
\end{align}
where we have denoted the phase differences as
$\varphi_{jk} = \theta_{j} - \theta_{k}$ and $\varphi_{j\ell} = \theta_{j} - \theta_{\ell}$.
Let us observe that $\Gamma_{1}$ is a $2\pi$-periodic function satisfying $\Gamma_{1}(\varphi_{jk} + 2\pi) = \Gamma_{1}(\varphi_{jk})$, and
$\Gamma_{2}$ is $2\pi$-periodic with respect to each argument, i.e., 
$\Gamma_{2}(\varphi_{jk} + 2\pi, \varphi_{j\ell}) = \Gamma_{2}(\varphi_{jk}, \varphi_{j\ell})$ and $\Gamma_{2}(\varphi_{jk}, \varphi_{j\ell} + 2\pi) = \Gamma_{2}(\varphi_{jk}, \varphi_{j\ell})$.

\subsection{Interaction functions yielding higher-order Kuramoto models}

Let us now make the key assumption that the pairwise interaction function $\tilde{\BG}_{1}$ takes a special functional form, given by 
\begin{align}
\label{eq:G1}
\tilde{\BG}_{1}(\BX_{j}, \BX_{k}) = \frac{1}{C} \BZ(\Theta(\BX_j)) \sin( \Theta(\BX_k) - \Theta(\BX_j) + \alpha ),
\end{align}
where $\alpha$ is a phase lag parameter for the pairwise PCF. 
Note that this interaction function is artificial in the sense that it explicitly depends on the phase values of the oscillators, which are non-physical quantities. 
However, such a coupling can also be realized from an engineering viewpoint, as discussed in Sec.~\ref{sec:conclusions}.
That is, we assume that the input $\tilde{\BG}_{1}(\BX_{j}, \BX_{k})$ from the oscillator $k$ to the oscillator $j$ is given in the direction of the PSF $\BZ(\Theta(\BX_j))$ of the $j$th oscillator, with the intensity being simply proportional to the phase coupling function that we aim to realize.
This pairwise interaction function gives an optimal pairwise IFP $\BG_{1}(\theta_{j}, \theta_{k}) = \frac{1}{C} \BZ(\theta_{j}) \sin(\theta_{k} - \theta_{j} + \alpha)$ that minimizes the averaged power of $\BG_{1}(\psi, \psi - \varphi_{jk})$ for each value of $\varphi_{jk}$ while realizing the sinusoidal PCF. 
We note that optimization is performed on the IFP, whose derivation is provided in Appendix~\ref{sec:appendix}.
Let us note that the value $C=\average{\norm{\BZ(\psi)}^{2}}{\psi}$ is constant, where $\norm{\cdot}$ denotes the Euclidean norm.
Therefore, the pairwise PCF is given by 
\begin{align}
\Gamma_{1}(\varphi_{jk}) &= \frac{1}{C} \average{\inner{\BZ(\psi)}{\BZ(\psi) \sin(-\varphi_{jk} + \alpha)}}{\psi} \cr
&= \sin(-\varphi_{jk} + \alpha),
\end{align}
which yields the following sinusoidal coupling function:
\begin{align}
\Gamma_{1}(\theta_{j} - \theta_{k}) &= \sin(\theta_{k} - \theta_{j} + \alpha).
\end{align}

We next assume two types of three-body interactions naturally derived from Stuart--Landau oscillators with all-to-all pairwise and higher-order interactions~\cite{Leon2024higher,Leon2025theory}.
In the same way as Eq.~\eqref{eq:G1}, the first type of three-body interaction functions is given by
\begin{align}
\label{eq:G2_Leon}
\tilde{\BG}_{2}(\BX_{j}, \BX_{k}, \BX_{\ell}) = \frac{1}{C} \BZ(\Theta(\BX_{j})) \sin(\Theta(\BX_{k}) + \Theta(\BX_{\ell}) - 2 \Theta(\BX_{j}) + \beta),
\end{align}
whose corresponding three-body PCF is then
\begin{align}
\Gamma_{2}(\varphi_{jk}, \varphi_{j\ell}) 
&= \frac{1}{C} \average{\inner{\BZ(\psi)}{\BZ(\psi) \sin(-\varphi_{jk} - \varphi_{j\ell} + \beta)}}{\psi} \cr
&= \sin(-\varphi_{jk} - \varphi_{j\ell} + \beta),
\end{align}
where $\beta$ is a phase lag parameter.
This three-body PCF yields the (1,1,-2)-sinusoidal coupling function 
\begin{align}
\Gamma_{2}(\theta_{j} - \theta_{k}, \theta_{j} - \theta_{\ell}) &= \sin(\theta_{k} + \theta_{\ell} - 2\theta_{j} + \beta).
\end{align}
The second type of three-body interaction functions is given by 
\begin{align}
\label{eq:G2_SA}
\tilde{\BG}_{2}(\BX_{j}, \BX_{k}, \BX_{\ell}) = \frac{1}{C} \BZ(\Theta(\BX_{j})) \sin(2\Theta(\BX_{k}) - \Theta(\BX_{\ell}) - \Theta(\BX_{j}) + \delta),
\end{align}
whose corresponding three-body PCF is then 
\begin{align}
\Gamma_{2}(\varphi_{jk}, \varphi_{j\ell}) 
&= \frac{1}{C} \average{\inner{\BZ(\psi)}{\BZ(\psi) \sin(-2\varphi_{jk} + \varphi_{j\ell} + \delta)}}{\psi} \cr
&= \sin(-2\varphi_{jk} + \varphi_{j\ell} + \delta),
\end{align}
where $\delta$ is a phase lag parameter.
This three-body PCF yields the (2,-1,-1)-sinusoidal coupling function 
\begin{align}
\Gamma_{2}(\theta_{j} - \theta_{k}, \theta_{j} - \theta_{\ell}) &= \sin(2\theta_{k} - \theta_{\ell} - \theta_{j} + \delta).
\end{align}
For clarity of notations, we denote the functions of $\sin(\theta_{k} + \theta_{\ell} - 2\theta_{j} + \beta)$ and $\sin(2\theta_{k} - \theta_{\ell} - \theta_{j} + \delta)$ as (1,1,-2)- and (2,-1,-1)-PCF, respectively, where the names are taken from the coefficients of $(\theta_{k}, \theta_{\ell}, \theta_{j})$.
These three-body interaction functions give optimal three-body IFPs that minimize the average power of $\BG_{2}(\psi,\psi - \varphi_{jk},\psi - \varphi_{j\ell})$ for each $(\varphi_{jk}, \varphi_{j\ell})$ while yielding the objective PCF.
The derivations of the optimal three-body IFPs are described in Appendix~\ref{sec:appendix}, where we also show that we can straightforwardly construct the optimal interaction functions also for fourth- or higher-order interactions.

Hence, the dynamics of each phase $j$ is given by
\begin{align}
\label{eq:Kuramoto_model_Leon}
\dot{\theta}_{j} ={}& \omega_{j} + \frac{K_{1}}{N} \sum_{k=1}^{N} \sin(\theta_{k} - \theta_{j} + \alpha) 
+ \frac{K_{2}}{N^{2}} \sum_{k,\ell=1}^{N} \sin(\theta_{k} + \theta_{\ell} - 2\theta_{j} + \beta)
\end{align}
when the three-body interaction function is given by the the form of Eq.~\eqref{eq:G2_Leon}, or given by
\begin{align}
\label{eq:Kuramoto_model_SA}
\dot{\theta}_{j} ={}& \omega_{j} + \frac{K_{1}}{N} \sum_{k=1}^{N} \sin(\theta_{k} - \theta_{j} + \alpha) 
+ \frac{K_{2}}{N^{2}} \sum_{k,\ell=1}^{N} \sin(2\theta_{k} - \theta_{\ell} - \theta_{j} + \delta)
\end{align}
when the three-body interaction function is given by the form of Eq.~\eqref{eq:G2_SA}.
We call such models higher-order Kuramoto models in this study.
The models~\eqref{eq:Kuramoto_model_Leon} and \eqref{eq:Kuramoto_model_SA} are two versions of the higher-order Kuramoto models obtained via phase reduction, whose difference arises from the interaction of the non-reduced model~\cite{Leon2025theory}. 
Both models have been thoroughly analyzed by Le\'{o}n \textit{et al.}~\cite{Leon2024higher} and by Skardal and Arenas~\cite{Skardal2020higher}, respectively.

Thus, the higher-order Kuramoto models with purely sinusoidal pairwise and three-body coupling functions can be derived from systems of arbitrary limit-cycle oscillators of nearly identical properties with pairwise and three-body all-to-all interactions, even though the limit cycle and PSF generally possess high-harmonic components.
If the three-body interactions are neglected (i.e., $K_{2} = 0$), both models are equivalent to the well-known Kuramoto--Sakaguchi model~\cite{Sakaguchi1986soluble}, as their pairwise interactions are the same.


\section{Analyses of the higher-order Kuramoto model}
\label{sec:analysis}

\subsection{Mean-field descriptions}

For the analysis of the higher-order Kuramoto models, we introduce the Kuramoto order parameter~\cite{Kuramoto1984chemical} defined by 
\begin{align}
\label{eq:order_parameter}
R\exp(i\Psi) = \frac{1}{N} \sum_{j=1}^{N} \exp(i\theta_{j}),
\end{align}
where $\Psi$ represents the collective phase, $R \in [0,1]$ is the order parameter that represents the amplitude of the collective oscillation, and $i = \sqrt{-1}$ is the imaginary unit. 
The order parameter takes $R = 0$ when the oscillators are in the completely incoherent state (i.e., no synchronization) and $R = 1$ when the system is in the fully synchronized state.
By using $R$ and $\Psi$, which represent the mean field of the system, the higher-order Kuramoto models~\eqref{eq:Kuramoto_model_Leon} and \eqref{eq:Kuramoto_model_SA} can be rewritten as
\begin{align}
\dot{\theta}_{j} ={}& \omega_{j} + K_{1} R \sin(\Psi - \theta_{j} + \alpha) + K_{2} R^{2} \sin(2\Psi - 2\theta_{j} + \beta)
\end{align}
and 
\begin{align}
\dot{\theta}_{j} ={}& \omega_{j} + K_{1} R \sin(\Psi - \theta_{j} + \alpha) + K_{2} R^{3} \sin(\Psi - \theta_{j} + \delta),
\end{align}
respectively.
The oscillator dynamics~\eqref{eq:oscillator_model} with the choices of the pairwise interaction function~\eqref{eq:G1} and three-body interaction function~\eqref{eq:G2_Leon} or \eqref{eq:G2_SA}, can also be rewritten as
\begin{align}
\label{eq:oscillator_model_Leon_simulation}
\dot{\BX}_{j} ={}& \BF_{j}(\BX_{j}) + \frac{K_{1}}{C} \BZ(\Theta(\BX_{j})) R \sin(\Psi - \Theta(\BX_{j}) + \alpha) 
+ \frac{K_{2}}{C} \BZ(\Theta(\BX_{j})) R^{2} \sin(2\Psi - 2\Theta(\BX_{j}) + \beta)
\end{align}
or
\begin{align}
\label{eq:oscillator_model_SA_simulation}
\dot{\BX}_{j} ={}& \BF_{j}(\BX_{j}) + \frac{K_{1}}{C} \BZ(\Theta(\BX_{j})) R \sin(\Psi - \Theta(\BX_{j}) + \alpha) 
+ \frac{K_{2}}{C} \BZ(\Theta(\BX_{j})) R^{3} \sin(\Psi - \Theta(\BX_{j}) + \delta),
\end{align}
respectively.
In our numerical study, we will use the formulation of Eqs.~\eqref{eq:oscillator_model_Leon_simulation} and \eqref{eq:oscillator_model_SA_simulation} for efficient computation.

\subsection{Collective phase control via Ott--Antonsen ansatz}

The higher-order Kuramoto dynamics with the (2,-1,-1)-PCF can be described solely by $R$ and $\Psi$ via the Ott--Antonsen~(OA) ansatz~\cite{Ott2008low}
when the three-body interaction function is given by the form of Eq.~\eqref{eq:G2_SA}.
If the phase lag parameters are $\alpha = 0$ and $\delta = 0$, the dynamics of $R$ and $\Psi$ are given by~\cite{Skardal2020higher}
\begin{gather}
\label{eq:OA_R}
\dot{R} = -\gamma R + \frac{K_{1}}{2} R(1 - R^{2}) + \frac{K_{2}}{2} R^{3}(1 - R^{2}), \\
\dot{\Psi} = \omega_{0},
\end{gather}
where the natural frequencies are assumed to follow the Lorentzian distribution with a width $\gamma$:
\begin{align}
P_{\mathrm{L}}(\omega) = \frac{1}{\pi}\frac{\gamma}{(\omega - \omega_{0})^{2} + \gamma^{2}}.
\end{align}
The dynamics~\eqref{eq:OA_R} has a stable solution~\cite{Skardal2020higher}:
\begin{gather}
R(t) = R_{0} = \sqrt{\frac{K_{2} - K_{1} + \sqrt{(K_{1} + K_{2})^{2} - 8\gamma K_{2}}}{2K_{2}}}, \\
\Psi(t) = \omega_{0} t + \mathrm{const.}
\end{gather}
in an appropriate range of the parameters $K_1$, $K_2$, and $\gamma$.

We now consider the oscillator model~\eqref{eq:oscillator_model_SA_simulation} without phase lags and assume that it receives a sufficiently small control input $\tilde{\Bp}_{j}(t) \in \mathbb{R}^{M}$ of order $\varepsilon$, described by
\begin{align}
\label{eq:oscillator_model_control}
\dot{\BX}_{j}(t) ={}& \BF_{j}(\BX_{j}(t)) + \frac{K_{1}}{C} \BZ(\Theta(\BX_{j}(t))) R(t) \sin(\Psi(t) - \Theta(\BX_{j}(t))) 
+ \frac{K_{2}}{C} \BZ(\Theta(\BX_{j}(t))) R(t)^{3} \sin(\Psi(t) - \Theta(\BX_{j}(t))) + \tilde{\Bp}_{j}(t).
\end{align}
We further assume that the control input to the phase model after phase reduction can be chosen as the product of $\sin(\theta_{j})$ and $u(t)$, i.e. $\inner{\BZ(\theta_{j}(t))}{\tilde{\Bp}_{j}(t)} = \sin(\theta_{j}(t))u(t)$.
The phase dynamics can then be described by
\begin{align}
\label{eq:Kuramoto_model_control}
\dot{\theta}_{j}(t) &= \omega_{j} + (K_{1}R(t) + K_{2}R(t)^{3})\sin(\Psi(t) - \theta_{j}(t)) + \inner{\BZ(\theta_{j}(t))}{\tilde{\Bp}_{j}(t)} \cr
&= \omega_{j} + (K_{1}R(t) + K_{2}R(t)^{3})\sin(\Psi(t) - \theta_{j}(t)) + \sin(\theta_{j}(t))u(t),
\end{align}
where $\sin(\theta_{j})$ and $u(t) \in \mathbb{R}$ are regarded as the individual PSF and a common control input for all the phase oscillators, respectively.
When the components of the PSF $\BZ$ are of order $1$, the common control input $u$ is of order $\varepsilon$.
The dynamics with the common control input~\eqref{eq:Kuramoto_model_control} can be reduced to the two-dimensional dynamical system via the OA ansatz~\cite{Fujii2025optimal}, which is described by
\begin{gather}
\label{eq:OA_R_control}
\dot{R}(t) = -\gamma R(t) + \frac{K_{1}}{2} R(t)(1 - R(t)^{2}) + \frac{K_{2}}{2} R(t)^{3}(1 - R(t)^{2}) - \frac{u(t)}{2} (1 - R(t)^{2})\sin(\Psi(t)), \\
\label{eq:OA_Psi_control}
\dot{\Psi}(t) = \omega_{0} + \frac{u(t)}{2} \left( \frac{1 + R(t)^{2}}{R(t)} \right)\sin(\Psi(t)).
\end{gather}


\section{Results}
\label{sec:results}


\subsection{FitzHugh--Nagumo oscillator}

As an example, we consider a system of  limit-cycle oscillators with pairwise and three-body all-to-all interactions described by Eq.~\eqref{eq:oscillator_model}, 
where the oscillators are given by the FitzHugh--Nagumo~(FHN) model~\cite{Fitzhugh1961impulses,Nagumo1962Active}, a paradigmatic model in mathematical neuroscience for the study of excitable systems and fast-slow oscillations~\cite{kuehn2015multiple,zhu2020phase}. 
The state $\BX = [x~y]^{\top}$ of the FHN model is driven by the vector field
\begin{align}
\label{eq:FHN}
\BF_{\mathrm{FHN}}(\BX) = 
\begin{bmatrix}
x - a x^{3} - y \\
c(x + b)
\end{bmatrix}
,
\end{align}
where $(a,b,c)$ are real-valued parameters.
The above vector field can admit a stable limit cycle and, hereafter, we call the FHN model with a stable limit cycle the FHN oscillator. 
When $c$ is small, this represents a slow-fast system and the timescale is determined by the parameter $c$, while the nullclines are independent of $c$.
The limit cycles and PSF are obtained numerically; note that, even if the oscillator models are simple, the limit cycle and PSF cannot be obtained analytically except in a few special cases~\cite{izhikevich2007dynamical,Nakao2016phase,leon2023analytical}.

\begin{figure}[t]
\centering
\includegraphics[width=0.6\textwidth]{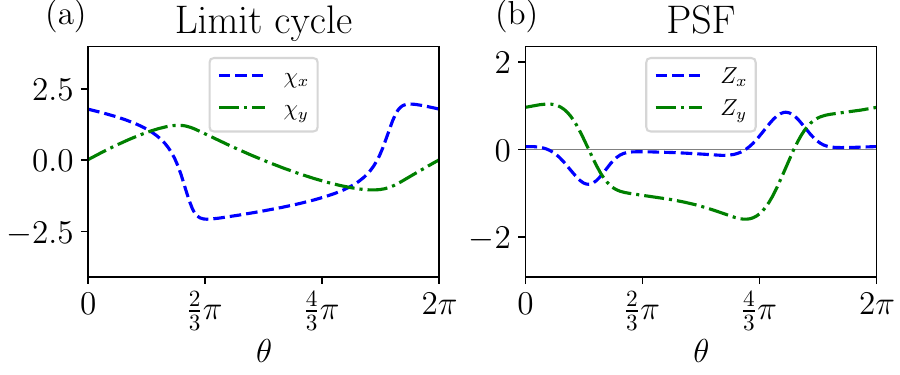}
\caption{
(a)~The limit-cycle orbit $\Bchi(\theta) = \left[ \chi_{x}(\theta)\; \chi_{y}(\theta) \right]^{\top}$ of the vector field~$\BF$ of the FHN oscillator as a function of $\theta$.
(b)~The PSF $\BZ(\theta) = \left[ Z_{x}(\theta)\; Z_{y}(\theta) \right]^{\top}$ of the vector field~$\BF$ of the FHN oscillator as a function of $\theta$.
}
\label{fig1}
\end{figure}

We consider a set of parameters $(a, b, c) = (1/3, 0.25, 0.15)$, where the vector field has an exponentially stable limit cycle. 
The vector field with these parameters is regarded as the common part $\BF$. 
In this setting, the limit-cycle oscillations have a period $T = 22.0$ and natural frequency $\omega_{0} = 0.2856$.
The limit-cycle orbit and PSF of the FHN oscillator are shown in Figs.~\ref{fig1}(a) and (b), respectively,
where we can observe that the functional forms are far from purely sinusoidal waves.

For calculating the interaction functions, we need to obtain the phase values, i.e., asymptotic phase function, of the FHN oscillators.
We can evaluate the phase values using an approximate mapping function from the two-dimensional state variable to an asymptotic phase, which is numerically evaluated before the simulation by running the FHN model from different initial conditions and measuring the asymptotic phase values~\cite{Nakao2016phase}. 
If the oscillator state $\BX$ is sufficiently close to the state on the limit cycle $\Bchi(\theta_{\mathrm{LC}})$ with a phase $\theta_{\mathrm{LC}}$, we can approximately evaluate an asymptotic phase value $\theta = \Theta(\BX)$ simply using the PSF~$\BZ$ by $\theta = \theta_{\mathrm{LC}} + \inner{\BZ(\theta_{\mathrm{LC}})}{(\BX - \Bchi(\theta_{\mathrm{LC}}))}$.

\subsection{Pairwise interactions: comparison with the classical Kuramoto model}

We first consider only pairwise interactions, i.e., $K_{2} = 0$.
In the pairwise Kuramoto model with the phase lag $\alpha = 0$, it is known that the transition from incoherent states to collective synchronization occurs when the pairwise coupling strength $K_{1}$ exceeds the critical value $K_{\mathrm{c}} = 2/(\pi P(0))$ in the $N \to \infty$ limit,
where $P(\omega)$ is a symmetric, smooth, and unimodal probability density function of the natural frequencies~\cite{Kuramoto1984chemical}.
Thanks to the chosen coupling, the system of the FHN oscillators exhibits the behavior predicted by the Kuramoto model, as we show numerically in what follows.

\begin{figure}[t]
\centering
\includegraphics[width=0.6\textwidth]{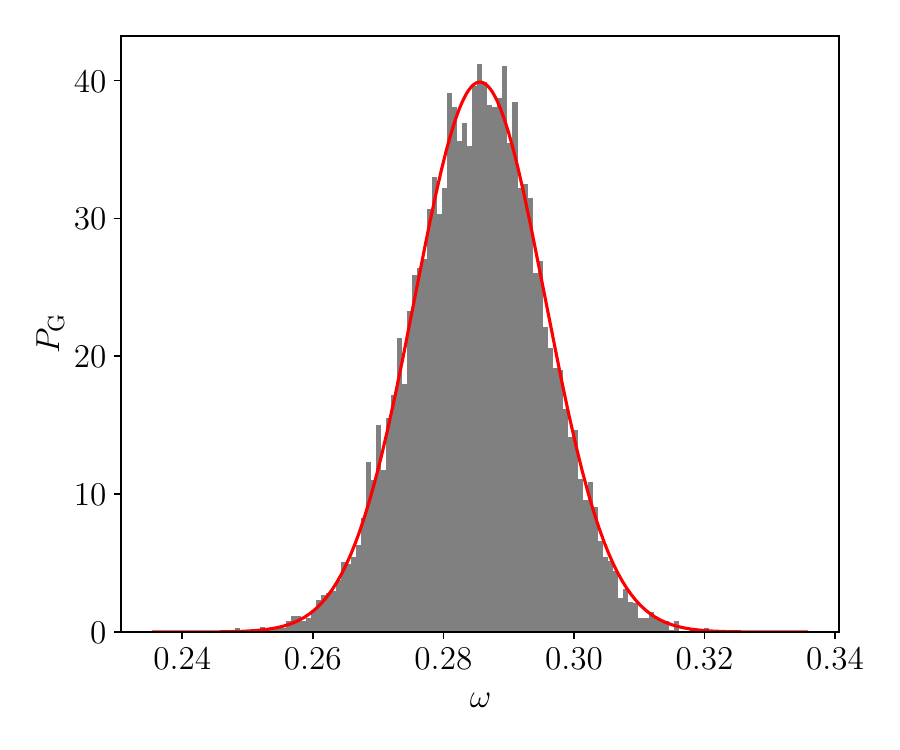}
\caption{
Comparison between the distribution $P_{\mathrm{G}}(\omega) = \calN(\omega;\omega_{0},\sigma^{2})$ and the normalized histogram of the natural frequencies of the FHN oscillators with the number of $N = 10^{4}$,
which are represented as the red curve and gray bars, respectively.
}
\label{fig2}
\end{figure}

We consider $N = 10^{4}$ FHN oscillators, whose frequency density $P$ follows the normal (Gaussian) distribution $P_{\mathrm{G}}(\omega) = \calN(\omega;\mu,\sigma^{2})$, where the mean $\mu$ is given by the natural frequency $\omega_{0}$ and the standard deviation is given by $\sigma = 0.01$.
Note that all the FHN oscillators obey the vector field $\BF_{\mathrm{FHN}}$, where the parameter $c$ of each oscillator is varied to yield a stable limit cycle with different natural frequencies.
The frequency distribution and normalized histogram of the natural frequencies of the FHN oscillators are depicted in Fig.~\ref{fig2}.
We numerically simulated the pairwise Kuramoto model from an incoherent initial condition and, analogously, the system of FHN oscillators with pairwise interactions from initial states on the limit cycle corresponding to an incoherent state of the Kuramoto model. 
For both cases, we used the fourth-order Runge--Kutta~(RK4) method with a time step $\Delta t = 10^{-1}$.
We computed the order parameters of both systems for $41$ different coupling strengths: $K_{1} = 0, 0.001, \dots, 0.04$. 

\begin{figure}[t]
\centering
\includegraphics[width=0.6\textwidth]{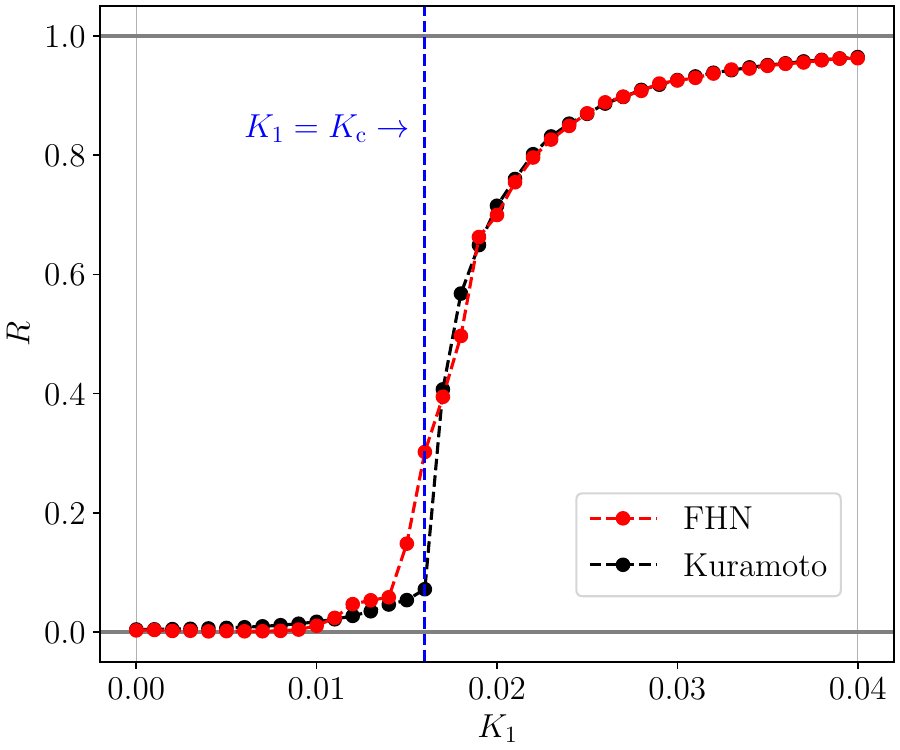}
\caption{
The order parameter $R$ vs.~the pairwise coupling strength $K_{1}$.
The red and black plots represent the order parameters of the system of the FHN oscillators and pairwise Kuramoto model, respectively.
The blue dotted vertical line represents the critical value $K_{1} = K_{\mathrm{c}}$.
We can observe that the system of the FHN oscillators also exhibits the synchronization transition when $K_{1}$ exceeds $K_{\mathrm{c}}$.
}
\label{fig3}
\end{figure}

\begin{figure}[t]
\centering
\includegraphics[width=0.6\textwidth]{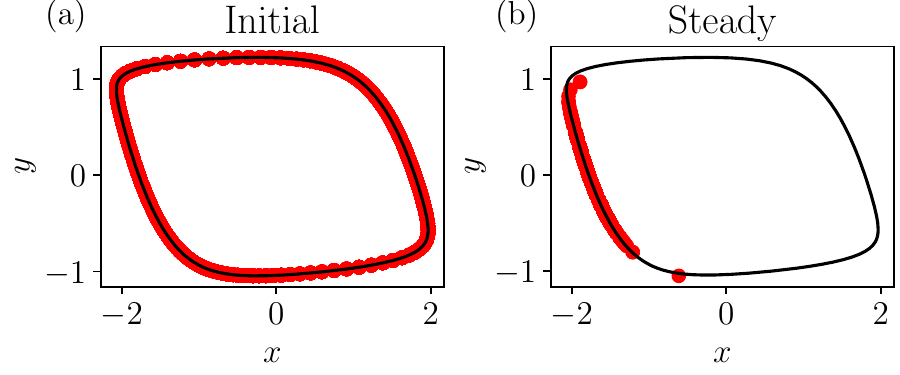}
\caption{
Distributions of the coupled FHN oscillators on the $xy$ plane when $K_{1} = 0.04$.
(a)~Incoherent initial state.
(b)~Collectively synchronized steady state.
In each panel, the red dots represent the oscillator states and the black periodic orbit represents the limit cycle of the FHN oscillator with the common part $\BF$.
}
\label{fig4}
\end{figure}

The results are shown in Fig.~\ref{fig3}, where we can observe that 
both systems exhibit collective synchronization transitions in which the value of $R$ sharply increases around $K_{1} = K_{\mathrm{c}}$. 
Note that the discrepancies between the two systems exist due to errors in the phase approximation, particularly near $K_{1} = K_{\mathrm{c}}$.
Since the peak value of a Gaussian distribution is proportional to $1/\sigma$, the critical value $K_{\mathrm{c}}$ decreases with $\sigma$.
Therefore, the transition can occur under weak interactions if the standard deviation of the natural frequencies is small enough, e.g., when the mean natural frequency is $\omega_{0} = 0.2856$ and standard deviation is $\sigma = 0.01$ as depicted in Fig.~\ref{fig2}.
Then, the critical value is $K_{\mathrm{c}} = 4\sigma / \sqrt{2\pi} \simeq 0.16$, as it can be observed in Fig.~\ref{fig3}.
In Figs.~\ref{fig4}(a) and (b), we show the initial and steady (i.e., after a sufficiently long transient) states of the system of coupled FHN oscillators on the $xy$ plane when $K_{1} = 0.04$.
We can observe that the oscillator states are distributed along the limit cycle in the initial state, whereas they are collectively synchronized near the limit cycle in the steady state. 

\subsection{Three-body interactions: comparison with the higher-order Kuramoto model}

Hereafter, we consider not only pairwise interactions, but also higher-order (three-body) interactions.
The three-body interactions for the (1,1,-2)-PCF can induce various dynamics such as clustering and slow-switching phenomena that cannot occur in the original Kuramoto model, as shown in the analysis by Le\'{o}n \textit{et al.}~\cite{Leon2024higher}.

\subsubsection{Anomalous transitions to synchrony}

Here, we assume that all FHN oscillators are identical and share the common vector field $\BF$ given by Eq.~\eqref{eq:FHN} with the above parameters.
For the higher-order Kuramoto model with identical frequencies, it is known that an anomalous transition to full synchrony occur in the following order: completely incoherent state, slow switching, two-cluster state, and fully synchronized state,
as the phase lag $\alpha$ of the pairwise coupling increases from $-\pi$ to $0$, e.g., when $K_{1} = 0.04$, $K_{2} = -0.018$ (i.e., $K_{2} / K_{1} = -0.45$), and $\beta = 0$ are assumed~\cite{Leon2024higher}.

\begin{figure}[t]
\centering
\includegraphics[width=0.7\textwidth]{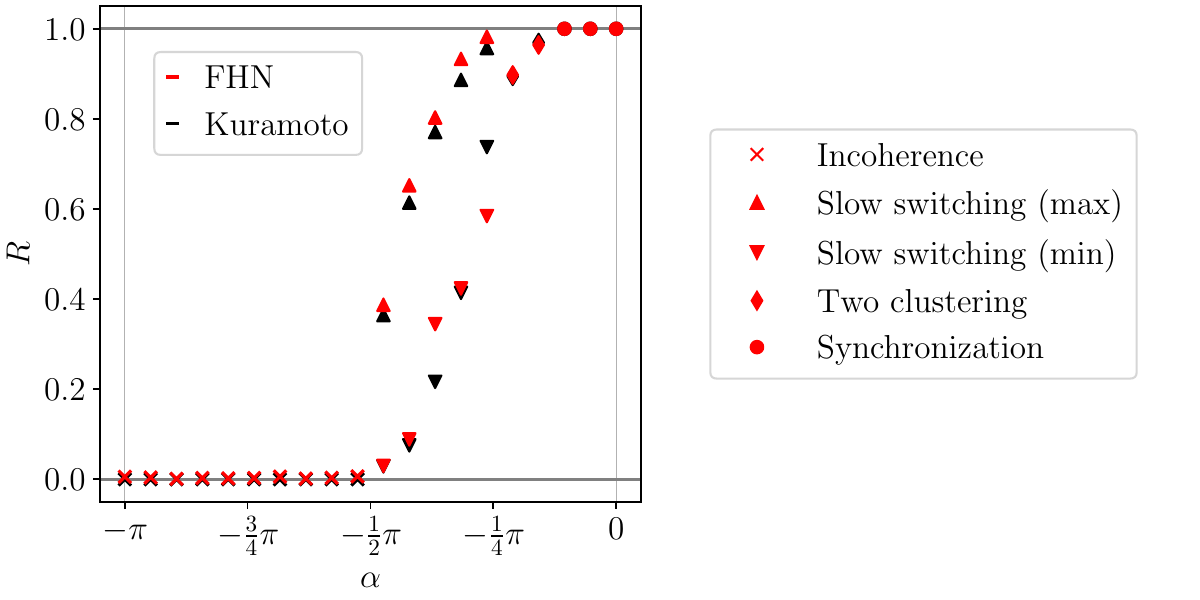}
\caption{
The order parameter $R$ vs.~the phase lag parameter $\alpha$.
The stable completely incoherent states, two-cluster states, and fully synchronized states are indicated by crosses, diamonds, and circles, respectively.
For slow-switching phenomena, the maximum and minimum values of the order parameters after the destabilization are indicated by upward and downward triangles, respectively. 
The red and black marks represent the results by the system of the FHN oscillators and higher-order Kuramoto model, respectively. 
}
\label{fig5}
\end{figure}

We numerically simulated the system of identical $N = 10^{3}$ FHN oscillators and the Kuramoto model with identical frequencies, both with pairwise and three-body interactions, for $20$ values of the phase lag $\alpha = -m\pi/19\; (m = 0, 1, \dots, 19)$ by the RK4 method with a time step $\Delta t = 10^{-1}$.
We then computed the order parameters $R$ for both models, which are shown in Fig.~\ref{fig5}.
For $\alpha \in [-\pi, -\pi/2)$, the system of FHN oscillators exhibits a stable incoherent state.
As $\alpha$ is increased, the incoherent state is destabilized at $\alpha \simeq -\pi/2$ and slow-switching phenomena occur up to $\alpha \lesssim -\pi/4$. 
The fully synchronized state is observed for $-\pi/8 \lesssim \alpha$.

Since the order parameter does not converge to a stationary value in the slow-switching state~\cite{Hansel1993clustering,Kori2001slow}, we compared the maximum and minimum values of the order parameter after the destabilization of the two-cluster state in this parameter regime.
There are differences in the maximum and minimum values of the order parameter between the two models, which arise from the inherent instability of slow-switching phenomena, that is, the saddle structures enlarge the tiny errors between the two models arising from phase approximation~\cite{Hansel1993clustering,Kori2001slow}.

We have thus shown that the same anomalous transition to synchrony observed in the higher-order Kuramoto model occurs also in the system of the FHN oscillators with the proposed interaction functions.
Note that this anomalous transition occurs even under weak interactions, because it depends on the ratio of $K_{2}$ to $K_{1}$, rather than on the intensity of $K_{1}$ and $K_{2}$, as reported in the previous work~\cite{Leon2024higher}. 
In fact, the FHN system being an example, we can claim that this occurs for any system of smooth limit cycle oscillators, as long as 
the interaction functions are given as described in Eq.~\eqref{eq:oscillator_model_Leon_simulation}.

\subsubsection{Two-clustering phenomena for systems with heterogeneous frequencies}

As another example, we focus on the two-clustering phenomenon in the higher-order Kuramoto model with identical frequencies
that arises when $K_{1} = 0.05$, $K_{2} = -0.06$ (i.e., $K_{2}/K_{1} = -1.2$), $\alpha = 0$, and $\beta = 1$ are assumed~\cite{Leon2024higher}. 
We demonstrate that the higher-order Kuramoto model and the FHN oscillators with the proposed interaction functions show two-clustering phenomena even under some heterogeneity in the oscillators.

Let us hence consider the case of heterogeneous FHN oscillators, i.e., the vector field for each oscillator $j$ is given by $\BF_{j}(\BX) = \BF(\BX) + \Bf_{j}(\BX)$ with different $\Bf_{j}(\BX)$. 
As in the higher-order Kuramoto model, we show that stable two-cluster states can be found also in the system of heterogeneous FHN oscillators, where each oscillator obeys the vector field $\BF_{\mathrm{FHN}}$ with a different parameter $c$.
We consider $N = 10^{4}$ FHN oscillators such that their natural frequencies follow a Gaussian distribution with the mean $\mu = \omega_{0}$ and standard deviation $\sigma = 0.01$, where the frequency distribution and normalized histogram of the natural frequencies are shown in Fig.~\ref{fig2}.

\begin{figure}[t]
\centering
\includegraphics[width=0.6\textwidth]{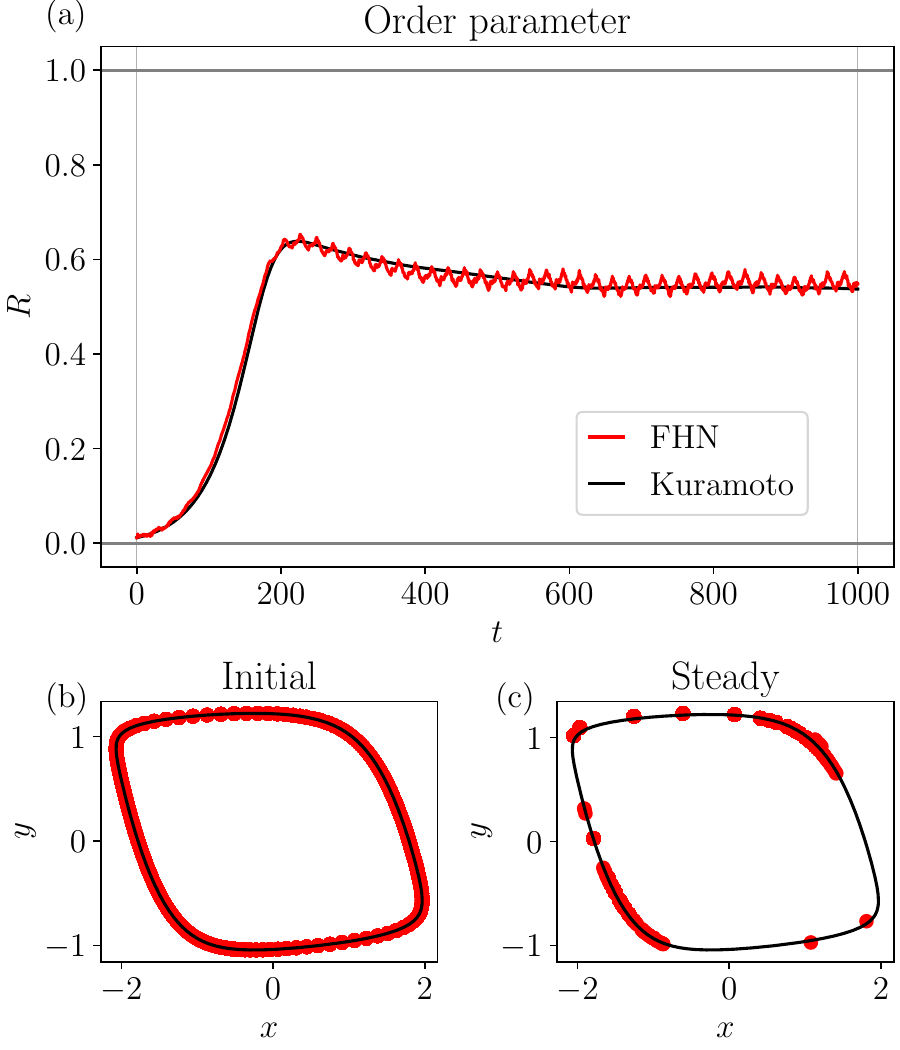}
\caption{
(a)~Comparison of the order parameter dynamics between the system of the FHN oscillators and higher-order Kuramoto model.
The incoherent initial state converges to the two-cluster state over time.
The red and black curves represent the dynamics of the system of FHN oscillators and higher-order Kuramoto model, respectively.
Since no temporal averaging is performed on the phases, the order parameter $R$ of the FHN oscillators exhibits small periodic fluctuations.
(b), (c)~Distributions of the states of the FHN oscillators on the $xy$ plane.
(b)~Incoherent initial state.
(c)~Steady two-cluster state.
In (b) and (c), the red dots represent the oscillator states and the black periodic orbit represents the limit cycle of the FHN oscillator with the common part $\BF$.
}
\label{fig6}
\end{figure}

\begin{figure}[t]
\centering
\includegraphics[width=0.6\textwidth]{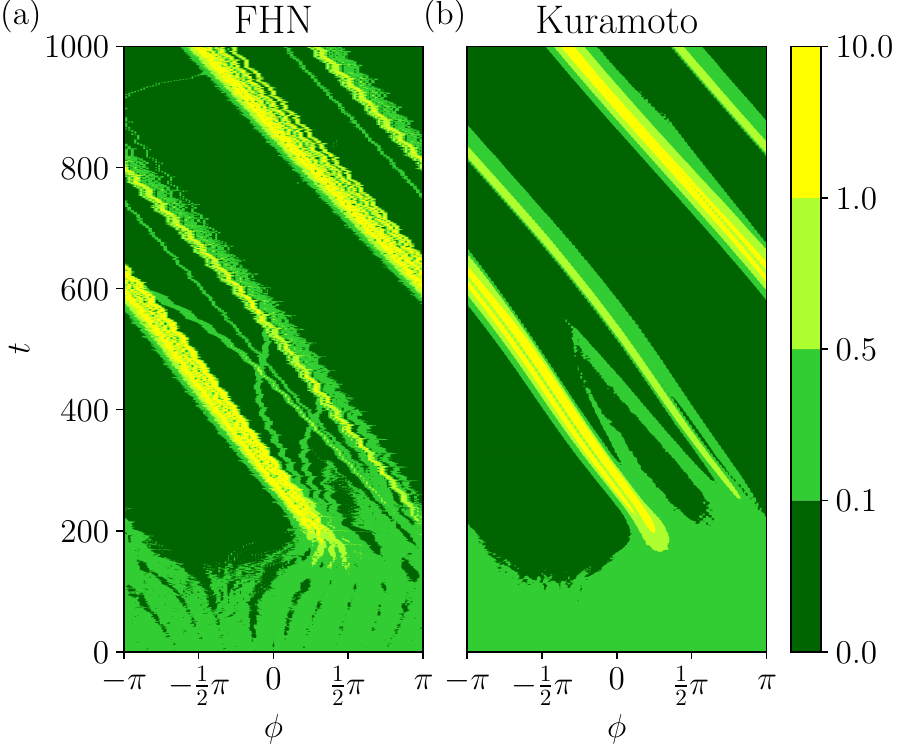}
\caption{
Comparison of the distributions (normalized histograms) of the relative phase between the system of FHN oscillators and higher-order Kuramoto model.
(a)~System of FHN oscillators.
(b)~Higher-order Kuramoto model.
}
\label{fig7}
\end{figure}

We numerically simulated the system of FHN oscillators and the higher-order Kuramoto model by the RK4 method with a time step $\Delta t = 10^{-1}$.
We then compare the order parameter dynamics of both models in Fig.~\ref{fig6}(a), and we find that both order parameters follow almost the same dynamics and converge sufficiently close to the stationary value around $t = 400$.
Note that the fluctuations in the order parameter of the FHN oscillators are mainly due to the averaging approximation errors.
The oscillator states in the initial state and in the steady state on the $xy$ plane are shown in Figs.~\ref{fig6}(b) and (c), respectively.
We can observe that the incoherent state along the limit cycle eventually exhibit a two-cluster state near the limit cycle.
Note that, due to the heterogeneity in the oscillators, they do not form strictly point clusters but rather distributed around them.
We then compare the dynamics of the normalized histograms of the relative phases $\phi_{j} = \theta_{j} - \omega_{0} t$ for the system of the FHN oscillators and higher-order Kuramoto model in Fig.~\ref{fig7}.
We find a good agreement in the time evolution of the distributions from the initial state into the two-cluster state, where both systems reach their steady states around $t = 400$.

Thus, the system of the FHN oscillators with the proposed interaction functions exhibits a two-clustering phenomenon even in the presence of heterogeneity, behaving similarly to the higher-order Kuramoto model. 

\subsection{Collective phase control via Ott--Antonsen ansatz}

As a final example, we control the collective phase of the FHN oscillators with the pairwise and three-body interaction functions for the (2,-1,-1)-PCF via the OA ansatz~\cite{Ott2008low,Skardal2020higher,Fujii2025optimal}.
We consider $N = 10^{4}$ FHN oscillators with different values of $c$, whose natural frequencies follow a Lorentzian distribution with a width $\gamma = 0.0001$.
We assume that the coupling strengths are $K_{1} = 0.004$ and $K_{2} = 0.002$, which yield the stable stationary value of the order parameter $R_{0} = 0.9830$.
We further assume that the common input is taken as $u(t) = -0.006 \times \cos(\omega_{0} t)$.
As mentioned in the analysis of OA control, we can take the control input to each FHN oscillator $\tilde{\Bp}_{j}(t) = [\tilde{p}_{j,x}(t)~\tilde{p}_{j,y}(t)]^{\top}$ so that it satisfies $\BZ_{x}(\theta_{j}(t))\tilde{p}_{j,x}(t) + \BZ_{y}(\theta_{j}(t))\tilde{p}_{j,y}(t) = \sin(\theta_{j}(t))u(t)$,
because not all components of the PSF simultaneously become zero as shown in Fig.~\ref{fig1}(b).
The relative collective phase $\Phi = \Psi - \omega_{0} t$ is expected to converge to $\Phi = 0$ because the time derivative $\dot{\Phi}$ satisfies $\dot{\Phi} \propto -\sin(\Phi)$ with the input $u(t)$, as predicted by averaging approximation of Eq.~\eqref{eq:OA_Psi_control} over one period $T$.

\begin{figure}[t]
\centering
\includegraphics[width=0.7\textwidth]{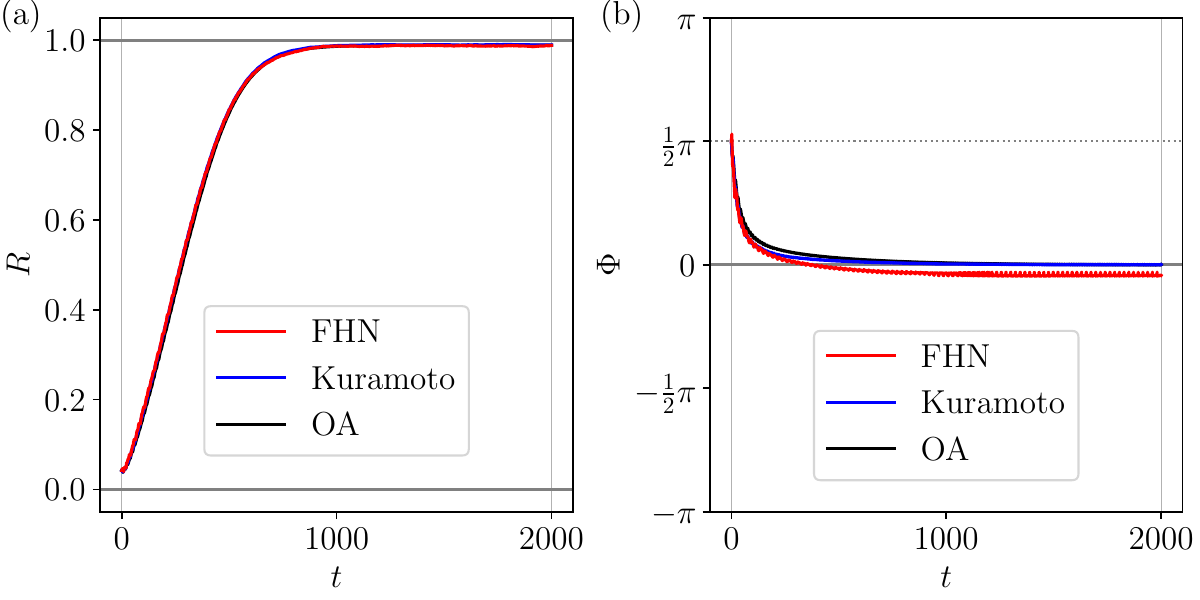}
\caption{
Comparison of the dynamics of the order parameter $R$ and the relative collective phase $\Phi = \Psi - \omega_{0} t$ for the system of FHN oscillators, higher-order Kuramoto model, and reduced OA system.
(a)~Order parameter dynamics.
(b)~Relative collective phase dynamics.
The red, blue, and black plots represent the dynamics of the system of the FHN oscillators, higher-order Kuramoto model, and reduced OA system, respectively.
In (b), the dotted line indicates the stationary behavior $\dot{\Phi} = 0$ when no control input is applied.
}
\label{fig8}
\end{figure}

We numerically simulated the system of FHN oscillators~\eqref{eq:oscillator_model_control}, higher-order Kuramoto model~\eqref{eq:Kuramoto_model_control}, and reduced OA system~\eqref{eq:OA_R_control},~\eqref{eq:OA_Psi_control} by the RK4 method with a time step $\Delta t = 10^{-1}$.
We compare the dynamics of the order parameter $R$ and relative collective phase $\Phi$ for the models in Figs.~\ref{fig8}(a) and (b), respectively. 
We find that the all the order parameters evolve from the incoherent state to the synchronized state almost in the same way and the relative collective phase follow almost the same dynamics starting from the initial value around $\Phi = \pi/2$ to the stationary value $\Phi = 0$, though there is a slight discrepancy in the relative collective phase of the FHN oscillators from $\Phi = 0$. 

Thus, the collective phase of the FHN oscillators can be controlled via the OA ansatz.
Note that such a control of the FHN oscillators based on the OA ansatz is possible thanks to the proposed interaction functions, and is not possible for simple interactions, e.g., linear diffusive interactions, which do not yield the Kuramoto-type phase models when reduced.


\section{Conclusions}
\label{sec:conclusions}

In this study, we designed optimal interaction functions between limit-cycle oscillators, from which pairwise and higher-order Kuramoto models can be derived exactly via phase reduction.
We numerically tested the behavior of a system of FHN oscillators with the derived interaction functions:
first, in the case only with pairwise interactions, by showing that it exhibited the synchronization transition when the coupling strength exceeded a critical value, as in the pairwise Kuramoto model; 
then, in the presence of three-body interactions for the (1,1,-2)-PCF, where we obtained the same anomalous transition to synchrony~\cite{Leon2024higher} as the higher-order Kuramoto model.
Moreover, we showed that the two-clustering phenomena occurred also in the system of slightly heterogeneous FHN oscillators, indicating that the characteristic behavior as predicted by the higher-order Kuramoto model occurs even when the properties of the oscillators are slightly different.
In addition, we demonstrated that we could control the collective phase of the FHN oscillators interacting through the pairwise and three-body interactions for the (2,-1,-1)-PCF via the OA ansatz.
Thus, the higher-order Kuramoto dynamics can be realized by the proposed optimal interaction functions.

The proposed interaction functions are made possible by the peculiar form of of the interaction functions chosen in Eqs.~\eqref{eq:G1}, \eqref{eq:G2_Leon}, and~\eqref{eq:G2_SA}, that is, they depend explicitly on the asymptotic phase values $\theta_{j, k, \ell} = \Theta(\BX_{j, k, \ell})$ of the oscillators rather than their original states $\BX_{j, k, \ell}$.
Let us note that this is an unusual assumption because the asymptotic phase is not a real physical quantity; 
usually, the interaction functions arise from real physical models of interacting limit-cycle oscillators,
like the linear diffusive coupling $\BG_{1}(\BX_{k}, \BX_{j}) \propto \BX_{k} - \BX_{j}$.
The proposed interaction functions depend explicitly on the asymptotic phase values, which require highly nonlinear transformation of the original variables.
Still, in principle, we can measure the asymptotic phase along the limit cycle, and we can also estimate the asymptotic phase from time-series data by several methods, such as the Hilbert transform~\cite{Pikovsky2001synchronization,gengel2021phase,matsuki2023extended}, polynomial fitting~\cite{namura2022estimating}, Gaussian process~\cite{yamamoto2025gaussian}, and neural networks~\cite{yawata2024phase,yawata2025data,yawata2025phase}. 
We think it feasible to compute the asymptotic phases and use them explicitly in the interaction functions in real time, considering the development of phase-estimation methods and efficiency of modern computers and hardwares such as field-programmable gate arrays.

The higher-order Kuramoto models can also be realized by injecting the interaction functions~\eqref{eq:G1}, \eqref{eq:G2_Leon}, and \eqref{eq:G2_SA} 
in the tangential direction along the limit cycle~\cite{yawata2025phase} rather than in the direction of the PSF, though it is not optimal under the present criterion on the power of the IFP.
In this study, we considered the interactions through complete state variables in the formulation, but we can also realize the higher-order Kuramoto models only by interacting partial state variables, where the asymptotic phase can be estimated from partial observations~\cite{yawata2025data}.
In addition, we considered up to three-body interactions, but the higher-order Kuramoto models can be derived in the same way when fourth- or higher-order interactions are considered, making the proposed framework widely applicable.
As is evident from the derivations in Appendix~\ref{sec:appendix}, the proposed method can straightforwardly be generalized to realize arbitrary pairwise and higher-order phase coupling functions among the oscillators, though we focused only on the simplest sinusoidal couplings of the fundamental harmonic.
Note also that the optimal interaction functions are not restricted to all-to-all interactions.

The optimal interaction functions can be important for synchronization engineering~\cite{kori2008synchronization}, where the interactions among the oscillators are designed to obtain a desired behavior. 
Moreover, the proposed framework can adopt artificially designed oscillators~\cite{Isjpeert2013dynamical,Ajalooeian2013general,Namura2023design,Namura2024designing}.
This study can be further extended by considering, for instance, stochastic systems~\cite{zhu2022phase,zhu2025complex}, edge dynamics~\cite{millan2020explosive,carletti2023global}, or more complex oscillators whose equations cannot be obtained analytically, involving both pairwise and higher-order interactions~\cite{Nakao2018phase,muolo2025synchronization}.


\section*{Acknowledgments}
We thank I. Le\'{o}n for useful discussions.
N. Namura acknowledges the financial support from JSPS KAKENHI 25KJ1270. 
The work of R. Muolo is supported by a JSPS postdoctoral fellowship, grant 24KF0211. 
H. Nakao acknowledges the financial support from JSPS KAKENHI 25H01468, 25K03081, and 22H00516.


\appendix

\section{Derivation of the optimal interaction functions}
\label{sec:appendix}

We derive the optimal interaction functions as functions of phases (IFPs) that yield the sinusoidal PCFs in the higher-order Kuramoto models.
We first derive the optimal pairwise interaction function $\BG_{1}$ that yields the sinusoidal PCF.
We can introduce the following representation for the IFP $\BG_{1}$:
\begin{align}
\BG_{1}^{\varphi_{jk}}(\psi) &= \BG_{1}(\psi,\psi - \varphi_{jk}),
\end{align}
where $\BG_{1}^{\varphi_{jk}}$ is $2\pi$-periodic with respect to $\psi$.
We can independently optimize the functional form of $\BG_{1}^{\varphi_{jk}}$ for each $\varphi_{jk}$, 
because $\BG_{1}^{\varphi_{jk}}$ and $\BG_{1}^{\tilde{\varphi}_{jk}}$ have no overlap with each other when $\varphi_{jk} \neq \tilde{\varphi}_{jk}$.

We seek the optimal $\BG_{1}^{\varphi_{jk}}$ that minimizes the averaged power for $\psi$ while the pairwise PCF $\Gamma_{1}$ is sinusoidal for each $\varphi_{jk}$.
We can denote the pairwise PCF to be realized as $h_{1}(\varphi_{jk})$ in a general manner; in the present case, $h_{1}(\varphi_{jk}) = \sin( - \varphi_{jk} + \alpha)$.
This optimization problem can be formulated as follows:

\begin{align}
\begin{aligned}
&\min_{\BG_{1}^{\varphi_{jk}}} && \frac{1}{2} \average{\norm{\BG_{1}^{\varphi_{jk}}(\psi)}^{2}}{\psi}  \\
&\mathrm{s.t.} && \Gamma_{1}(\varphi_{jk}) = \average{\inner{\BZ(\psi)}{\BG_{1}^{\varphi_{jk}}(\psi)}}{\psi} = h_{1}(\varphi_{jk}). \\
\end{aligned}
\end{align}
This optimization problem can be analytically solved by the method of Lagrange multipliers.
We introduce a functional
\begin{align}
S\left\{ \BG_{1}^{\varphi_{jk}},\lambda_{1} \right\} = \frac{1}{2} \average{\norm{\BG_{1}^{\varphi_{jk}}(\psi)}^{2}}{\psi} + \lambda_{1} \left( \average{\inner{\BZ(\psi)}{\BG_{1}^{\varphi_{jk}}(\psi)}}{\psi} - h_{1}(\varphi_{jk}) \right),
\end{align}
where $\lambda_{1}$ is a Lagrange multiplier.
From the extremum condition for $S$ with respect to $\BG_{1}^{\varphi_{jk}}$:
\begin{align}
\fd{S}{\BG_{1}^{\varphi_{jk}}} &= \BG_{1}^{\varphi_{jk}}(\psi) + \lambda_{1} \BZ(\psi) = 0,
\end{align}
we obtain 
\begin{align}
\BG_{1}^{\varphi_{jk}}(\psi) &= -\lambda_{1}\BZ(\psi),
\end{align}
where we find that $\BG_{1}^{\varphi_{jk}}$ is proportional to the PSF $\BZ$.
Considering the constraint on the pairwise PCF, the Lagrange multiplier should satisfy 
\begin{align}
\lambda_{1} = -\frac{h_{1}(\varphi_{jk})}{\average{\norm{\BZ(\psi)}^{2}}{\psi}}.
\end{align}
Thus, the optimal functional form of $\BG_{1}^{\varphi_{jk}}$ is obtained as
\begin{align}
\label{eq:G1_optimal}
\BG_{1}^{\varphi_{jk}}(\psi) ={}& \BG_{1}(\psi,\psi - \varphi_{jk}) \cr
={}& \frac{1}{C} \BZ(\psi) h_{1}(\varphi_{jk}),
\end{align}
where $C = \average{\norm{\BZ(\psi)}^{2}}{\psi}$ is a constant.
We note that the optimal pairwise IFP~\eqref{eq:G1_optimal} is proportional to $\BZ(\psi)$.
When the PCF is given by the sinusoidal function, i.e., $h_{1}(\varphi_{jk}) = \sin(-\varphi_{jk} + \alpha)$, the optimal pairwise IFP is given as a product of the sinusoidal PCF and the waveform of the PSF $\BZ(\psi)$.
In the expression using the set of the phases $(\theta_{j},\theta_{k})$, the optimal pairwise IFP can be represented as 
\begin{align}
\BG_{1}(\theta_{j},\theta_{k}) = \frac{1}{C} \BZ(\theta_{j}) \sin(\theta_{k} - \theta_{j} + \alpha).
\end{align}
Note that the above derivation applies to any PCF $h_{1}(\varphi_{jk})$.

Next, we derive the optimal three-body IFP $\BG_{2}$ that yields the sinusoidal PCFs in a similar way.
We can introduce the following representation for the three-body IFP $\BG_{2}$:
\begin{align}
\BG_{2}^{\varphi_{jk},\varphi_{j\ell}}(\psi) &= \BG_{2}(\psi,\psi - \varphi_{jk},\psi - \varphi_{j\ell}),
\end{align}
where $\BG_{2}^{\varphi_{jk},\varphi_{j\ell}}$ is $2\pi$-periodic with respect to $\psi$.
We can independently optimize the functional form of $\BG_{2}^{\varphi_{jk},\varphi_{j\ell}}$ for each $(\varphi_{jk},\varphi_{j\ell})$, 
because $\BG_{2}^{\varphi_{jk},\varphi_{j\ell}}$ and $\BG_{2}^{\tilde{\varphi}_{jk},\tilde{\varphi}_{j\ell}}$ have no overlap with each other when $(\varphi_{jk},\varphi_{j\ell}) \neq (\tilde{\varphi}_{jk},\tilde{\varphi}_{j\ell})$.

We seek the optimal $\BG_{2}^{\varphi_{jk},\varphi_{j\ell}}$ that minimizes the averaged power for $\psi$ while the three-body PCF $\Gamma_{2}$ is sinusoidal for each $(\varphi_{jk},\varphi_{j\ell})$.
We can also formulate the optimization problem for a general PCF $h_{2}(\varphi_{jk},\varphi_{j\ell})$:
\begin{align}
\begin{aligned}
&\min_{\BG_{2}^{\varphi_{jk},\varphi_{j\ell}}} && \frac{1}{2} \average{\norm{\BG_{2}^{\varphi_{jk},\varphi_{j\ell}}(\psi)}^{2}}{\psi}  \\
&\mathrm{s.t.} && \Gamma_{2}(\varphi_{jk},\varphi_{j\ell}) = \average{\inner{\BZ(\psi)}{\BG_{2}^{\varphi_{jk},\varphi_{j\ell}}(\psi)}}{\psi} = h_{2}(\varphi_{jk}, \varphi_{j\ell}).
\end{aligned}
\end{align}
This optimization problem can also be solved analytically.
We introduce a functional
\begin{align}
S\left\{ \BG_{2}^{\varphi_{jk},\varphi_{j\ell}},\lambda_{2} \right\} = \frac{1}{2} \average{\norm{\BG_{2}^{\varphi_{jk},\varphi_{j\ell}}(\psi)}^{2}}{\psi} + \lambda_{2} \left( \average{\inner{\BZ(\psi)}{\BG_{2}^{\varphi_{jk},\varphi_{j\ell}}(\psi)}}{\psi} - h_{2}(\varphi_{jk}, \varphi_{j\ell}) \right),
\end{align}
where $\lambda_{2}$ is a Lagrange multiplier.
From the extremum condition with respect to $\BG_{2}^{\varphi_{jk},\varphi_{j\ell}}$:
\begin{align}
\fd{S}{\BG_{2}^{\varphi_{jk},\varphi_{j\ell}}} &= \BG_{2}^{\varphi_{jk},\varphi_{j\ell}}(\psi) + \lambda_{2} \BZ(\psi) = 0,
\end{align}
we obtain 
\begin{align}
\BG_{2}^{\varphi_{jk},\varphi_{j\ell}}(\psi) &= -\lambda_{2}\BZ(\psi),
\end{align}
where we also find that $\BG_{2}^{\varphi_{jk},\varphi_{j\ell}}$ is proportional to the PSF $\BZ$.
Considering the constraint on the three-body PCF, the Lagrange multiplier should satisfy 
\begin{align}
\lambda_{2} = -\frac{h_{2}(\varphi_{jk}, \varphi_{j\ell})}{\average{\norm{\BZ(\psi)}^{2}}{\psi}}.
\end{align}
Thus, the optimal functional form of $\BG_{2}^{\varphi_{jk},\varphi_{j\ell}}$ is obtained as
\begin{align}
\label{eq:G2_optimal}
\BG_{2}^{\varphi_{jk},\varphi_{j\ell}}(\psi) ={}& \BG_{2}(\psi,\psi - \varphi_{jk},\psi - \varphi_{j\ell}) \cr
={}& \frac{1}{C} \BZ(\psi) h_{2}(\varphi_{jk}, \varphi_{j\ell}).
\end{align}
We note that the optimal three-body IFP~\eqref{eq:G2_optimal} is also proportional to $\BZ(\psi)$.
When the PCF is given by $h_{2}(\varphi_{jk}, \varphi_{j\ell}) = \sin(-\varphi_{jk} - \varphi_{j\ell} + \beta)$ or $h_{2}(\varphi_{jk}, \varphi_{j\ell}) = \sin(-2\varphi_{jk} + \varphi_{j\ell} + \delta)$,
the optimal three-body IFP can be represented as 
\begin{align}
\BG_{2}(\theta_{j},\theta_{k},\theta_{\ell}) = \frac{\BZ(\theta_{j}) \sin(\theta_{k} + \theta_{\ell} - 2\theta_{j} + \beta)}{\average{\norm{\BZ(\psi)}^{2}}{\psi}}
\end{align}
or
\begin{align}
\BG_{2}(\theta_{j},\theta_{k},\theta_{\ell}) = \frac{\BZ(\theta_{j}) \sin(2\theta_{k} - \theta_{\ell} - \theta_{j} + \delta)}{\average{\norm{\BZ(\psi)}^{2}}{\psi}},
\end{align}
respectively, in the expression using the set of the phases $(\theta_{j},\theta_{k},\theta_{\ell})$.

The fourth- or higher-order interaction functions can be derived in the same way if necessary. 
In addition, though we assumed the sinusoidal PCFs in this study, we can easily solve the optimization problems for the IFPs also for different functional forms of the PCFs.


\bibliography{references}

\end{document}